\documentclass[pre,amsmath,amssymb,twocolumn,superscriptaddress]{revtex4}

\usepackage{amsmath,amssymb}
\usepackage[usenames]{color}
\usepackage{amssymb}
\usepackage{dsfont}
\usepackage{grffile}
\usepackage[pdftex]{graphicx}
\usepackage{subfigure}
\usepackage{amsmath, amstext, amssymb, amsfonts, amsxtra, url}
\usepackage{textcomp}
\usepackage{xspace}
\DeclareSymbolFontAlphabet{\amsmathbb}{AMSb}

\newcommand{\uw}{\uparrow}
\newcommand{\dw}{\downarrow}
\newcommand{\Rec}{\mathcal{R}}
\newcommand{\Con}{\mathcal{C}}
\newcommand{\ket}[1]{|\!\! #1 \rangle}
\newcommand{\bra}[1]{\langle #1 \!\!|}

\newcommand{\rhop}{{\rho}}

\newcommand{\sop}{{\sigma}}
\newcommand{\Dop}{\mathcal{D}}

\newcommand{\An}{{A}_{L}(\omega)}
\newcommand{\Ad}{{A}_{L}^{\dagger}(\omega)}

\newcommand{\w}{\omega}
\newcommand{\la}{\lambda}

\newcommand{\kbc}{k_B}

\begin{document}

\title{Thermal Rectification in Classical and Quantum Systems: Searching for Efficient Thermal Diodes}

\author{Emmanuel Pereira}
\affiliation{Departamento de F\'{\i}sica--Instituto de Ci\^{e}ncias Exatas, Universidade Federal de Minas Gerais, CP 702, 30.161-970 Belo Horizonte MG, Brazil}

\begin{abstract}
This  mini-review addresses a bedrock problem for the advance of phononics: the building of feasible and efficient thermal diodes. We revisit investigations in classical and quantum systems. For the classical anharmonic chains of
oscillators, the most used model for the study of heat conduction in insulating solids, we recall the ubiquitous occurrence of thermal rectification in graded systems, and we show that the match between graded structures and
long range interactions is an efficient mechanism to increase the rectification factor. For the cases of genuine quantum models, we present the spin chains, such as the open $XXZ$ model, as profitable systems for the occurrence of
thermal rectification and other interesting related properties. In particular, we describe two cases of perfect diodes: one for the spin current, in a two-segmented $XXZ$ model, and another one for the heat current in a simple
quantum Ising model with long range interactions. We believe that such results involving interesting rectification properties in simple models will stimulate more theoretical and experimental investigations on the subject.

\end{abstract}


\maketitle

\section{Introduction}

Two preeminent physical mechanisms of transporting energy, especially from a practical viewpoint, are conduction by electricity and by heat. But their status in science are quite different. On one hand, we have
the amazing development of modern electricity that transformed our daily lives with the creation of technological apparatus, possible due to the invention of electrical diodes, transistors and other non-linear solid
states devices. On the other hand, however, considering the manipulation of heat currents and management of thermal circuits, the scenario is much poorer, consequence of the lacking of viable and powerful thermal diodes.

The thermal diode, or rectifier, is a system in which the magnitude of the heat flow changes as it is  inverted between two thermal baths, i.e., it is a system with a preferable direction for the heat current. The
thermal diode is the basic ingredient of the devices aiming the control of the heat flow. The phenomenon of thermal rectification was first observed by Starr in the 1930's \cite{S}, but the first manageable theoretical model
for the thermal diode was proposed only in 2002 \cite{Terraneo}. The most recurrent  proposals of diode, including this first one, are given by the sequential coupling of two or three segments of different classical anharmonic
oscillators. Unfortunately, they present critical problems: the rectification power is small and rapidly decays to zero as the system size increases.

This mini-review is focused on several attempts devoted to overcome this difficult problem, i.e., devoted to the building of a feasible thermal diode with a big rectification factor. Having in mind the disclosure of the ingredients
responsible for the phenomenon, we consider only ``simple'' classical and quantum models. However, it is worth to recall the existence of other intricate schemes (ignored here) involving geometry or complicate interactions.
For example, carbon nano-structures with elaborate shapes and asymmetries \cite{JL}, graphene nanoribons \cite{Hu, nano2}, etc. In Ref.\cite{BLi2}, a rectification factor of 350\% is reported in a graphene nanoribon with a quite
specific two-dimensional trapezium shape. More details about these elaborate proposal with geometrical nanostructures are found in Ref.\cite{RMPBLi}.

Most of the models used in the investigation of the microscopic mechanisms of heat current are given by classical dynamical systems. However, the possibility of effects of quantum nature in low temperatures, together with the
present ambient of device miniaturization, makes mandatory the study of energy flow in genuine quantum models. Therefore, we visit some recent works on quantum spin chains such as the $XXZ$ model, in which rectification ubiquitously holds. Our results include  cases of  perfect rectification: one of the spin current and another one of the heat flow.

The rest of this mini-review is organized as follows. In the second section we present some results involving the classical chains of oscillators. In the third section we turn to the genuine quantum models, here given by quantum spin chains. The last section is devoted to final remarks.

\section{Chains of classical oscillators}

Since the pioneering works of Debye \cite{D} and Peierls \cite{P}, the most recurrent models for the study of heat conduction in insulating solids are given by chains (lattices) of anharmonic oscillators, precisely, by systems of
$N$ oscillators with Hamiltonian such as
\begin{eqnarray}
H(q,p) &=& \sum_{j=1}^{N} ~ \frac{1}{2}\left [\frac{p_j^2}{m_{j}} ~ + ~ M_{j}q_j^2\right ]
+ \frac{1}{2}\sum_{j\neq l=1}^{N}q_lJ_{lj}q_j \nonumber \\
& & ~ + ~
\sum_{j=1}^{N}\lambda_{j} \mathcal{P}(q_j)
\end{eqnarray}
where $m_{j}$ is the mass of the j-th oscillator; $q$ and $p$ are vectors in $\mathbb{R}^{N}$, denoting displacement and momentum; $\lambda,  M_j >0$; $J_{jl}=J_{lj}=f(|\ell-j|)$, $f$ with some integrable decay;  ${P}$ is the anharmonic on-site potential, for
example, ${P}(q_j) = q_{j}^{4}/4$.  Excellent reviews on this and related problems are found in Refs. \cite{LLP, Dh1, RMPBLi}.
The dynamics of the systems involves the coupling with the baths and are usually given by the stochastic differential
equations
\begin{equation}
dq_j\! =\! {p_j} dt , ~~
dp_j\! =\! -\frac{\partial H}{\partial q_j}dt-\zeta_{j}
p_jdt+\gamma^{1/2}_j dB_j  , \label{eqdynamics}
\end{equation}
where the baths $B_j$ are independent Brownian motions, with zero average and diffusion equal to 1
\begin{equation} \label{noise}
\left< B_{j}(t)\right> = 0, ~~ \left< B_{j}(t)B_{\ell}(s)\right> = \delta_{i,j}{\rm min}(t,s) ,
\end{equation}
 $\zeta_{j}$ is the constant coupling between site
$j$ and its reservoir; $\gamma_j=2\zeta_j m_j T_j$, where
$T_j$ is the temperature of the $j$-th  bath. For the case of baths only at the boundaries we take $\zeta_j = 0$ for $j=2, \ldots, N-1$.

To study the heat flow, we write the total energy as the sum over the energy of each single oscillator, and then we analyze the continuity equation associated to this energy. Precisely, we write
\begin{equation}
H_{j}(q_{j},p_{j})\! = \!\ \frac{p_j^2}{2}   + \frac{1}{2}\sum_{l\neq j}V(q_l - q_j) + V_{2}(q_j)  , \label{Hamiltoniansingle}
\end{equation}
where $H(q,p) = \sum_{j}H_{j}(q_{j},p_{j})$, and
$V$ and $V_{2}$ follows from the previous definition for the Hamiltonian. From the dynamical equations we obtain
\begin{eqnarray}
\left<\frac{dH_j}{dt}(t) \right> &=&  \mathcal{R}_j(t) + \left< \mathcal{F}_{\rightarrow j}-\mathcal{F}_{j\rightarrow} \right>,
\\
\mathcal{F}_{j\rightarrow} &=& \sum_{\ell>j}\nabla_{j} V(q_j-q_{\ell})\left(\frac{p_j}{2}+\frac{p_{\ell}}{2}\right) \nonumber\\
&  = & \sum_{\ell>j}J_{j\ell}(q_j-q_{\ell})\left(\frac{p_j}{2}+\frac{p_{\ell}}{2}\right)\!\!,\label{fluxo1}
\\
\mathcal{F}_{\rightarrow j} &=& \sum_{\ell<j}\nabla_{j} V(q_j-q_{\ell})\left(\frac{p_j}{2}+\frac{p_{\ell}}{2}\right) \nonumber \\
& = & \sum_{\ell<j}J_{j\ell}(q_j-q_{\ell})\left(\frac{p_j}{2}+\frac{p_{\ell}}{2}\right)\!\!, \label{fluxo2}
\\
\mathcal{R}_j(t) &=& \zeta_{j}\left( T_{j} - \left< p_{j}^{2}\right>\right) \label{reservatorio}.
\end{eqnarray}
$\mathcal{F}_{j\rightarrow}$ denotes the heat current from site $j$ to the forward sites $\ell>j$; $\mathcal{F}_{\rightarrow j}$ describes the current
from the previous sites $\ell<j$. $\mathcal{R}_{j}$ gives the energy flux between the $j$-th site and the $j$-th reservoir: it is zero for $j= 2, \ldots N-1$ in the case of baths only in the boundaries, and it also vanishes for
the case of inner reservoirs in the self-consistent conditions (details ahead). The expressions show us that the heat current is given by simple two-point correlation functions. Anyway, the nonlinear dynamical equations for the case of
anharmonic potentials make exceedingly difficult their analytical study.

The easier harmonic case $\lambda_{j} \equiv 0$ has analytical solution, see the seminal paper of Rieder, Lebowitz and Lieb \cite{RLL}, and also the very interesting article by Casher and Lebowitz \cite{CL}.

Unfortunately, as well known, the pure harmonic chain of oscillators does not obey the Fourier's law: the regime of heat transport is ballistic, the flow does not decay with the system size. Moreover, there is no thermal rectification in any asymmetrical version of the harmonic chain. Thus, in the search for thermal rectification, we are forced to consider some anharmonicity in the models.

In this direction, the self-consistent harmonic chain is an old proposal of effective anharmonic model \cite{Bosterli}, recurrently revisited \cite{BLL, PF-PRE, Dh2}. It is given by the harmonic chain of oscillators with inner
stochastic reservoirs ($\zeta_{j} \neq 0$ also for $j =2, \ldots, N-1$ in Eq.(\ref{reservatorio})), whose temperatures, however, are chosen such that there is no net heat flow between an inner reservoir and its linked site in the
steady state. That is, the self-consistent condition guarantees that, in the steady state, the heat current across the chain is supplied only by the baths at the boundaries. The inner stochastic reservoirs describe only some
mechanism of phonon scattering, i.e., they are a footprint of the anharmonic potentials absent in the Hamiltonian. Such a trace of anharmonicity seems to be indeed present: in the harmonic chain of oscillators with self-consistent
inner baths, Fourier's law holds \cite{BLL, PF-PRE}, in contrast with the ballistic purely harmonic chain. However, in any asymmetric version of this model, there is no rectification \cite{PLA-PRE, Divira}.

At this stage, it worth to recall that, in Ref.\cite{P-2017}, we propose an effective harmonic classical model, with extra temperature dependent harmonic potentials, which represent, in some sense, the dynamical average
effect of the absent anharmonic potentials. Asymmetrical versions of the model rectify. It helps us to understand that a temperature dependence of some inner structure or mechanism is crucial for the phenomena occurrence, but the problem of a large rectification
factor is left open.

The most recurrent models of thermal diodes involve genuine anharmonic systems, precisely, they are given by the coupling of two or three segments with different anharmonic potentials \cite{Terraneo, LiCasati}. The rectification
phenomenon here is explained by the idea of phonon-band matching (miss-matching), but the small rectification factor together with a rapid vanishment with the system size increasement points out the necessity of different proposals.

In the search of general properties, i.e., solutions independent on the specificity of the model, an iteresting attempt appears with the use of graded materials, aiming the growth of asymmetry and, consequently, of the rectification in
the chain. Graded models are inhomogeneous systems whose structure changes gradually in space. They can be found in nature and can also be manufactured, having attracted interest in many areas \cite{Gr}: electricity, optics,
mechanics, etc.

The first work on rectification in graded systems \cite{YBLi} was carried out in a model with a Fermi-Pasta-Ulam $\beta$ potential and graded mass distribution, with $m_{j}$ always changing between two fixed values, $M_{max}$ and
$M_{min}$, which, in some sense, inhibits the growth of asymmetry as the size of the chain increases. The authors show the occurrence of thermal rectification and also the existence of regions with negative differential thermal resistance.

Analytical results indicating the ubiquity of thermal rectification in graded systems appear in Refs.\cite{Prapid2010, Psuf}. Sufficient conditions for the occurrence of the phenomenon are established \cite{Psuf}, and the possibility of enhancing the rectification as one increases the asymmetry together with the system size is pointed out. In particular, the ubiquity is made transparent with the on-set of rectification in a quite simple, naked
model of bars and graded bullets with elastic collisions \cite{WPC}, now by means of computer simulations.

Anyway, the establishment of conditions which guarantees the occurrence of thermal rectification in anharmonic graded chains does not suppress the necessity of a large rectification factor. In such a direction, some works appear
with ideas of general scope. Precisely, graded chains with long range interactions (LRI) are detailed studied in Refs.\cite{PA-PRE, CPC}, i.e., systems of $N$ oscillators with Hamiltonian such as
\begin{equation}
\mathcal{H} = \sum_{j=1}^{N} \left( \frac{p_{j}^{2}}{2m_{j}} + \frac{q_{J}^{4}}{4}\right) + \sum_{j,k} \frac{(q_{j} - q_{k})^{2}}{2(1 + |j - k|^{\lambda})} ~,
\end{equation}
where $q_{j}$ gives the displacement of the $j$th oscillator; $p_{j}$ its momentum; $m_{j}$ is the mass, which has a graded distribution; $\lambda$ is the power for the polynomial decay with distance of the interparticle interaction.

Here, the scenario is the following. First, we expect the occurrence of rectification with the anharmonicity and the graded distribution. Intuitively, the LRI favors the heat flow with the addition of new links, i.e., new channels
for the heat transport. More importantly, these new channels connect distant particles with very different masses. In other words, the LRI favors the asymmetry, which in turn favors the increase of rectification. Moreover, as
we increase the system size with the graded mass distribution, new particles are introduced, creating new asymmetric channels of heat transport due to the LRI. In short, we expect that, in a graded system with LRI, the thermal
rectification will be bigger and will not decay as the system size increases. Analytical investigations \cite{PA-PRE} and detailed numerical simulations confirm these statements \cite{CPC}. See Fig.1. There, the rectification factor versus the system size is depicted for systems with LRI and nearest-neighbor interaction. The rectification factor is defined as $f_{r}=\frac{(\mathcal{F}_{+}-\mathcal{F}_{-})}{\mathcal{F}_{-}}\times100\%$,
where $\mathcal{F}_{+}$ and $\mathcal{F}_{-}$ represent, respectively, the larger heat
flow and the smaller heat flow, which are measured by inverting the
temperatures of the baths at the two ends of the chain. It is worth to note that, for the case of the hot and cold temperatures given by $T_{H}=9.9$ and $T_{C}=0.1$, $N=64$, $m_{1}=1$, $m_{N}=10$, $\lambda = 1.2$, we obtain the significative value $f_{r}=4638\%$.

\begin{figure}[!]
\includegraphics[width=\columnwidth]{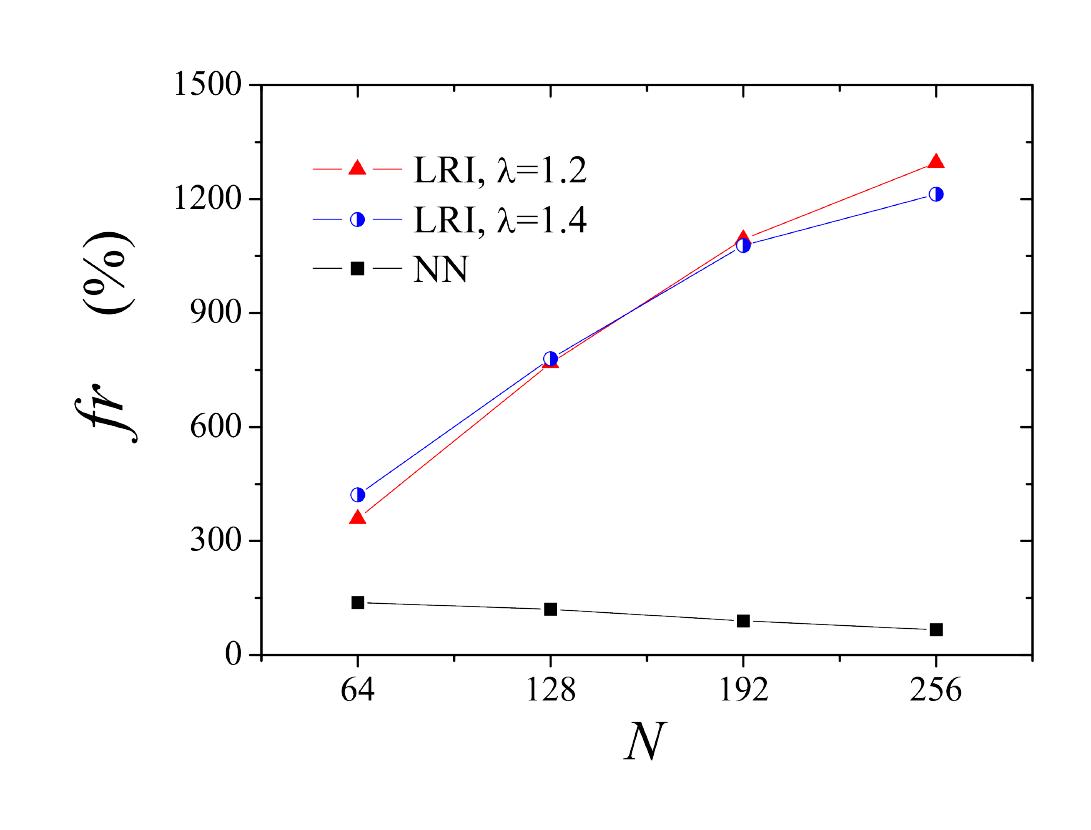}
\vskip-.4cm \caption{(Color on line) Dependence of rectification factor on the system size $N$. Here $T_{H} = 9.5$, $T_{C}= 0.5$, $m_{1}=1$. Triangles are for LRI with $\lambda = 1.2$, circles are for LRI with $\lambda = 1.4$, squares are for the NN case. The mass gradient is 1/64. ($m_N=2$ for N=64; $m_N=3$ for N=128; $m_N=4$ for N=192; $m_N =5$ for N=256.). The plots are borrowed from Ref.\cite{CPC}.}
\end{figure}

Some further comments are convenient here. Besides its role played in rectification, the investigation of LRI in heat transport is intricate and deserves attention by itself, see, e.g., Ref.\cite{Iubini} and references there in.
It is also interesting to note the proposal of mechanisms for the thermal rectification far away from complicate geometrical shapes and also from the graded distribution and LRI used here: in Ref.\cite{Muga}, for example, the
authors show the phenomenon occurrence in a homogeneous chain with a local modification of the interactions at one site only.

\section{Quantum spin chains}

The detailed investigation of genuine quantum models becomes a must, given the existence of quantum effects in the low temperature heat conduction. The study is also highly stimulated by the present ambient of
miniaturization in science due the advance of nanotechnology, lithography, etc.

In the context of systems of oscillators, we recall that the quantum harmonic chain with inner reservoirs in the self-consistent condition is examined in Refs.\cite{DRoy, PPLA, PLA-PRE, D2}. Interestingly, in opposition to the
behavior of the classical self-consistent harmonic chain, now, due to the quantum nature of the inner reservoir that brings temperature to the bulk of the system, there is thermal rectification in the quantum graded version
\cite{PPLA, PLA-PRE, D2}.

However, a more interesting class of models appears here: the quantum spin chains, such as the $XXZ$ open system. These models seems to be more treatable than the quantum anharmonic chains and, besides that, there is a huge interest in the study of such models by themselves: $XXZ$ is the archetypal model to the study of open quantum systems, with recurrent and increasing attention in several areas, such as optics, cold atoms, condensed matter, quantum
information \cite{T, BP}.

The one-dimensional, spin $1/2$ $XXZ$ model with $N$ sites, nearest-neighbor interactions and in the presence of an external magnetic field $B$ is given by the Hamiltonian
(for $\hbar = 1$)
\begin{eqnarray}
\mathcal{H} = && \sum_{i=1}^{N-1}\left\{ J_{i,i+1}\left( \sigma_{i}^{x}\sigma_{i+1}^{x} + \sigma_{i}^{y}\sigma_{i+1}^{y} \right) + \Delta_{i,i+1}\sigma_{i}^{z}\sigma_{i+1}^{z} \right\}\nonumber \\
 && + \sum_{i=1}^{N} B_{i}\sigma_{i}^{z} ~, \label{hamiltonian}
\end{eqnarray}
where $\sigma_{i}^{\delta}$ ($\delta = x, y, z$) are the Pauli matrices. The Markovian dynamics, in the case of boundary driven models with target polarization at the edges, is given by the Lindblad master equation (LME) for the
density matrix $\rho$ \cite{BP}
\begin{equation}
\frac{d\rho}{d t} = i[\rho, \mathcal{H}] + \mathcal{D}(\rho) ~.\label{master}
\end{equation}
In our first analysis, we take the dissipator $\mathcal{D}(\rho)$ as given by
\begin{eqnarray}
\mathcal{D}(\rho) &=& \mathcal{D}_{L}(\rho) + \mathcal{D}_{R}(\rho) ~, \nonumber\\
\mathcal{D}_{L,R}(\rho) &=& \sum_{s=\pm} L_{s}\rho L_{s}^{\dagger} -
\frac{1}{2}\left\{ L_{s}^{\dagger}L_{s} , \rho \right\} ~,\label{dissipator}
\end{eqnarray}
where, for $\mathcal{D}_{L}$, we have
\begin{equation}
L_{\pm} = \sqrt{\frac{\gamma}{2}(1 \pm f_{L})} \sigma_{1}^{\pm} ~\label{dissipator2},
\end{equation}
and a similarly for $\mathcal{D}_{R}$, but with $\sigma_{N}^{\pm}$ and $f_{R}$ replacing  $\sigma_{1}^{\pm}$ and $f_{L}$. In the formulas above, $\{\cdot,\cdot\}$ denotes the anticommutator;
$\sigma_{j}^{\pm}$ are the spin creation and annihilation operators $\sigma_{j}^{\pm} = (\sigma_{j}^{x} \pm i\sigma_{j}^{y})/2$~; $\gamma$ is the coupling strength to the spin baths; $f_{L}$ and $f_{R}$ are the
driving strength. In this first case considered here,  the baths are modeled in terms of extra spins $\sigma_{0}^{z}$  and $\sigma_{N+1}^{z}$ linked to the chain, and so, $f_{L}$ and $f_{R}$ describe
different spin polarization at the boundaries: precisely, $f_{L} = \left<\sigma_{0}^{z}\right>$ and  $f_{R} = \left<\sigma_{N+1}^{z}\right>$.

In what follows, we take homogeneous $J_{i,i+1} \equiv J$ and graded asymmetry parameter $\Delta_{i,i+1}$.

Turning to the currents for the quantities spin and energy, precise expressions can be derived from the continuity equation and from the LME for the dynamics. For the spin flow, we have
\begin{eqnarray}
\frac{ d\langle \sigma_{1}\rangle}{dt} &=& \langle \mathcal{J}_{L} \rangle - \langle \mathcal{J}_{1} \rangle ~, \nonumber \\
\frac{ d\langle \sigma_{i}\rangle}{dt} &=& \langle \mathcal{J}_{i-1} \rangle - \langle \mathcal{J}_{i} \rangle ~,  ~ 1< i < N ~, \nonumber \\
\frac{ d\langle \sigma_{N}\rangle}{dt} &=& \langle \mathcal{J}_{N-1} \rangle - \langle \mathcal{J}_{R} \rangle ~,
\end{eqnarray}
where, by setting $f_{L}= f = -f_{R}$,
\begin{eqnarray}
\langle \mathcal{J}_{j} \rangle &=& 2 J \langle \sigma_{j}^{x} \sigma_{j+1}^{y} - \sigma_{j}^{y}\sigma_{j+1}^{x} \rangle ~, ~2\leq j \leq N-1 ~, \nonumber \\
                     &=& 4i J \langle \sigma_{j}^{+} \sigma_{j+1}^{-} - \sigma_{j}^{-}\sigma_{j+1}^{+} \rangle ~,  \label{spinc} \\
\langle \mathcal{J}_{L} \rangle &=& \gamma \left( f - \langle \sigma_{1}^{z} \rangle \right) ~,  \nonumber \\
\langle \mathcal{J}_{R} \rangle &=& -\gamma \left( f + \langle \sigma_{N}^{z} \rangle \right) ~.
\end{eqnarray}

In the steady state we have a homogeneous flow through the chain, i.e.,
\begin{equation}
\langle \mathcal{J}_{1} \rangle_{S} = \langle \mathcal{J}_{2} \rangle_{S} = \ldots = \langle \mathcal{J}_{N} \rangle_{S} \equiv \langle \mathcal{J} \rangle ~.
\end{equation}

In order to derive the expression for the energy current, we first split the Hamiltonian equation as sum of bonds
\begin{eqnarray}
H &=&  \sum_{i=1}^{N-1} \varepsilon_{i,i+1} = \sum_{i=1}^{N-1} h_{i,i+1} + b_{i,i+1} ~,  \\
h_{i,i+1} &=& J \left( \sigma_{i}^{x} \sigma_{i+1}^{x} + \sigma_{i}^{y}\sigma_{i+1}^{y} \right) + \Delta_{i,i+1} \sigma_{i}^{z} \sigma_{i+1}^{z} ~, \nonumber\\
b_{i,i+1} &=& \frac{1}{2} \left[ B_{i}\sigma_{i}^{z}(1+\delta_{i,1}) + B_{i+1}\sigma_{i+1}^{z}(1+\delta_{i+1,N}) \right]~.\nonumber
\end{eqnarray}
Moreover, we separate above the term coming from the XXZ interaction from the part related to the external magnetic field. For the inner sites, $2\leq i\leq N-1$, we have
\begin{equation}
\frac{ d\langle \varepsilon_{i,i+1}\rangle}{dt} = \langle F_{i} \rangle - \langle F_{i+1} \rangle ~,
\end{equation}
and so, for the energy current,
\begin{equation}
\langle F_{j} \rangle = i\langle [\varepsilon_{j-1,j}, \varepsilon_{j,j+1}] \rangle ~, ~2\leq j \leq N-1 ~.
\end{equation}

It is also quite convenient to separate the energy current into two parts, namely,
\begin{equation}
\langle F_{i}\rangle = \langle F_{i}^{XXZ}\rangle + \langle F_{i}^{B}\rangle ~,
\end{equation}
where the XXZ contribution is (for $2\leq j\leq N-1$)
\begin{eqnarray}
\lefteqn{\langle F_{j}^{XXZ} \rangle = i\langle [h_{j-1,j}, h_{j,j+1}] \rangle = \ldots } \nonumber\\
&=& 2 J \langle J \left( \sigma_{j-1}^{y}\sigma_{j}^{z} \sigma_{j+1}^{x} - \sigma_{j-1}^{x}\sigma_{j}^{z}\sigma_{j+1}^{y}\right) \nonumber\\
&& + \Delta_{j-1,j}\left( \sigma_{j-1}^{z}\sigma_{j}^{x} \sigma_{j+1}^{y} - \sigma_{j-1}^{z}\sigma_{j}^{y}\sigma_{j+1}^{x}\right) \nonumber\\
&& + \Delta_{j,j+1}\left( \sigma_{j-1}^{x}\sigma_{j}^{y} \sigma_{j+1}^{z} - \sigma_{j-1}^{y}\sigma_{j}^{x}\sigma_{j+1}^{z}\right)\rangle  ~.\label{Jxxz}
\end{eqnarray}
And, again for $2\leq j\leq N-1$,
\begin{eqnarray}
\langle F_{j}^{B} \rangle &=& i\langle [\varepsilon_{j-1,j}, \varepsilon_{j,j+1}] - [h_{j-1,j}, h_{j,j+1}] \rangle = \ldots  \nonumber\\
 &=& \frac{1}{2} B_{j}\langle \mathcal{J}_{j-1} + \mathcal{J}_{j}\rangle ~,
 \end{eqnarray}
 where $\mathcal{J}_{j}$ is the spin current, already defined.

It is interesting to remark that, for the homogeneous XXZ chain in the presence of an uniform magnetic field $B$, we have $\langle F_{j}^{XXZ}\rangle = 0$ \cite{PopLi}, and so,
$\langle F \rangle = B \langle \mathcal{J} \rangle$, that is, the total energy current essentially coincides with the spin current. Moreover, for the case of a graded chain, it follows that
$\langle F \rangle = \langle F^{XXZ} \rangle + B \langle \mathcal{J} \rangle$, where, due to arguments of symmetry in the LME \cite{Prapid2017},  $\langle F^{XXZ} \rangle$ is a nonvanishing  even function of $f$, and $\langle \mathcal{J} \rangle$ is an odd function of $f$
(the spin current direction is always determined by the direction of the magnetization imbalance). Thus, the thermal rectification is transparent, i.e.,  the change in the magnitude of the energy current as we invert the baths ($f \leftrightarrow -f$): the energy current is given by the sum of an even and another odd function of $f$. Such ubiquity of the thermal rectification occurrence is confirmed by direct calculations, in different situations, of the currents in the steady state of the model \cite{SPG}. It is also immediate, in the case of a zero magnetic field $B$, the occurrence of the ``one-way street'' phenomenon, i.e., as $\langle F_{j}^{XXZ}\rangle$ is an even function of $f$, the direction of the energy current
in graded system will not change as we invert the baths. It will be determined only by the asymmetry in the graded chain. Recall that, in the homogeneous chain, we have $\langle F_{j}^{XXZ}\rangle = 0$, as said.
The thermodynamic consistency of this phenomenon is shown by splitting the energy current into heat and power (work), see details in Refs.\cite{PHeat2018, FBarra}.

In fact, as richer as the equilibrium $XXZ$ model, which presents ferro, antiferro, gapless phases, etc., here,  the open $XXZ$ graded chain gives us a privileged scenario. Let us give a further example, concerning the occurrence of thermal rectification and manipulations of the heat flow. By choosing the interaction parameters we can build a graded material such that we have rectification in one direction (e.g., the larger heat current flowing from the left to the
right side). Then, it is possible to get the reversal of rectification (i.e., larger flow from the right to the left side) by adding a proper external magnetic field, but keeping the material without changing its parameters
\cite{Alberto}. In a few words, we observe in the model the occurrence of themal rectification as well as the possibility of getting its reversal by means of an external magnetic field.

Another problem of interest is the investigation of two segmented $XXZ$ system, which imitates the first models of classical diodes, given by the coupling of different anharmonic segments. Instead of a graded system, we take a bipartite chain, both with the same $J$ (the $XY$ coupling), see
Eq.(\ref{hamiltonian}), but with
different $\Delta$, i.e., $\Delta_{j,j+1}= \Delta_{L}$ for $j< N/2$, and $\Delta_{j,j+1} = \Delta_{R}$ for the other half $j>N/2$. We couple the chain to two different baths at its edges, each one tending to impose a particular
magnetization to the system. Precisely, we assume the dynamics given by a LME with the dissipator $\mathcal{D} = \sum_{n=1,N} \mathcal{D}_{n}$, where
\begin{align}
\Dop_n(\rhop)=&\gamma \left[ \la_n \left(\sop^+_n \rhop \sop^-_n - 1/2 \left\{\sop^-_n\sop^+_n,\rhop\right\}  \right)   \right.   \nonumber \\
&+\left.  (1-\la_n) \left(      \sop^-_n \rhop \sop^+_n - 1/2 \left\{\sop^+_n\sop^-_n,\rhop\right\} \right)   \right].
\end{align}
The parameter $\gamma$ above gives the intensity of the coupling to the baths; the spin magnetization imposed by the baths is set by $\la_n$. We assume  $\la_1$ and $\la_N$ equal to $0$ or $0.5$, i.e., on one side of the chain the bath tends to set the spins to be pointing down ($\ket{\dw}_n\bra{\dw}$ for $\la_n=0$), or to be in an equal mixture of up and down spins ($[\ket{\dw}_n\bra{\dw} + \ket{\uw}_n\bra{\uw}]/2$ for $\la_n=0.5$).
 The imbalance $\lambda_1-\lambda_N$ imposed by the baths provokes a spin current.
For the case $\la_1>\la_N$, which we name as ``forward bias'', the spin current flows from left to right, and the opposite, the ``reverse bias'', corresponds to $\la_1<\la_N$, in which the current flows from right to left.

We use two quantities to measure the spin rectification. Namely, the coefficient $\mathcal{R} = - \mathcal{J}_{f}/\mathcal{J}_{r}$, and the contrast $\mathcal{C} =
|(\mathcal{J}_{f} +\mathcal{J}_{r})/(\mathcal{J}_{f} - \mathcal{J}_{r})|$. We have $\mathcal{R} = 1$ and $\mathcal{C}= 0$ for the absence of rectification, and for a perfect diode, $\mathcal{R}$ and $\mathcal{C}$ tend to infinity
and $1$ respectively. We perform a study taking $\Delta_{R}=0$ and different values for $\Delta_{L}/J_{L}$ \cite{PRL2018}. Surprisingly, and contrasting with the behavior of the classical chains of oscillators, we find a huge
spin rectification that increases with the system size, tending to a perfect diode, for $\Delta_{L}/J_{L} > 3$. See Fig.2.

\begin{figure}
\includegraphics[width=\columnwidth]{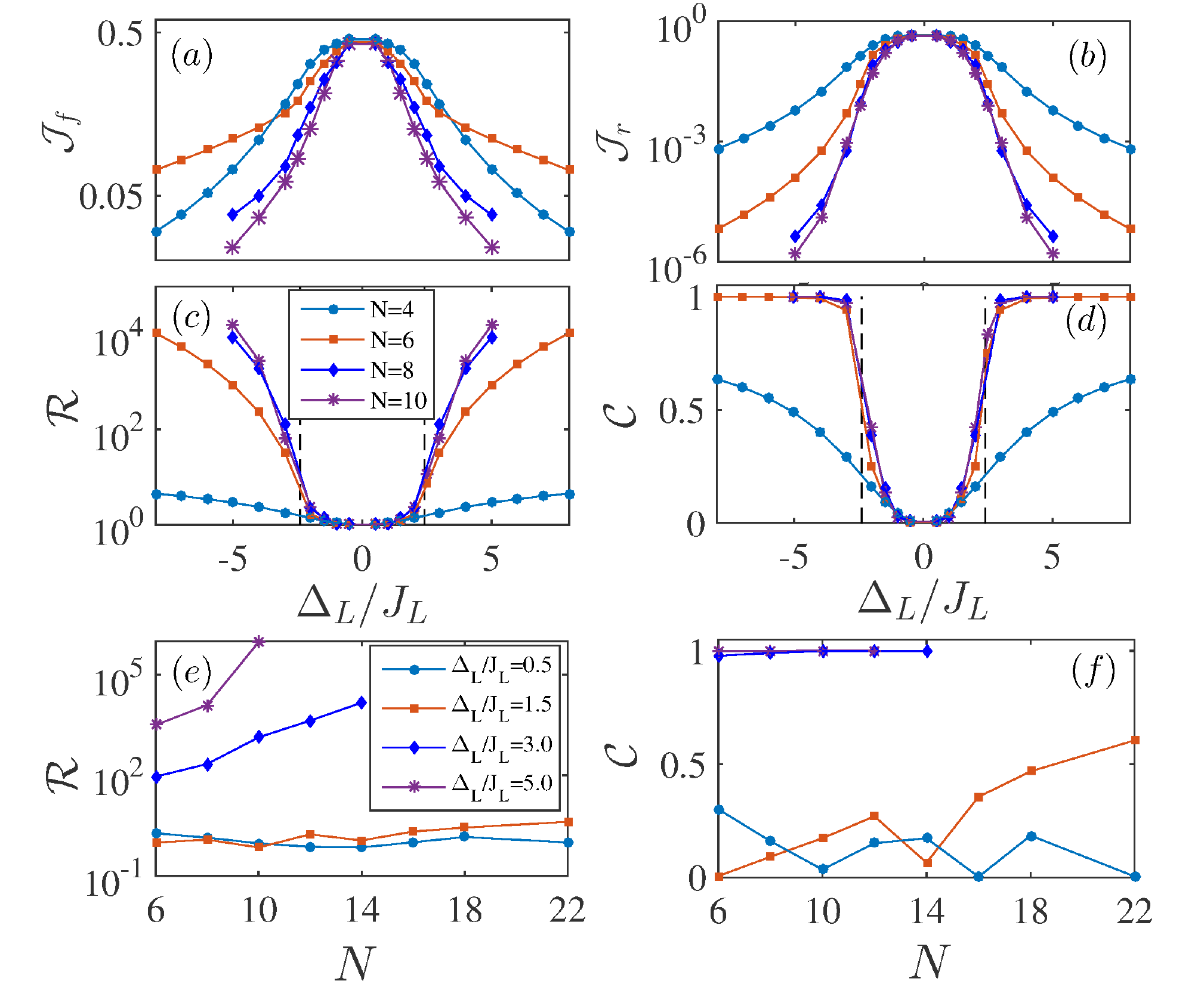}
\caption{(Color online) 
We plot, as a function of the anisotropy $\Delta_L/J_L$ for different chain lengths $N$, the forward (a), reverse currents (b), the rectification coefficient (c) and the contrast (d).
The current in reverse bias is significantly lower than in forward bias.
Rectification $\Rec$ (e) and contrast $\Con$ (f) as functions of the system size for different values of the anisotropy $\Delta_L/J_L$. The rectification $\Rec$ increases significantly with the system size for large enough anisotropy $\Delta_L/J_L$. At the junction between the two halves of the chain $\Delta_{N/2}=\Delta_{M}$ and $J_{N/2}=J_{M}$.
The other parameter values are $\Delta_R=\Delta_M=0$, $J_L=J_R$, $J_M=J_L$ and $\gamma=J_L/\hbar$ for (a-d), while $J_M=0.1 J_L$ and $\gamma=0.1 J_L/\hbar$ for (e-f). The plots are borrowed from Ref.\cite{PRL2018}. }
\label{fig:Fig1}
\end{figure}

For this two segmented $XXZ$ chain but with global baths acting on all eigenstates of the Hamiltonian system (details ahead), we study the heat rectification and show that it can be large if the interaction is strong enough in one
half of the chain and if one bath is at a cold enough temperature \cite{BBPCP}. In short, for the heat flow, we obtain a big rectification, but not a perfect diode.

To go further in the search for suitable thermal diodes, it seems profitable to keep studying spin chains, but possibly some simplified model instead of the intricate $XXZ$ chain. Discarding intricate effects and interactions
may be useful to understand the minimal ingredients behind the rectification phenomenon.

Thus, we turn to the quantum Ising model, say, a simple component of the Heisenberg ($XXZ$) family. Precisely, we take a quantum spin chain with Hamiltonian
\begin{equation}
H_{S} = \sum_{i=1}^{N} h_{i}\sigma_{i}^{z} + \sum_{i,k} \Delta_{i,k} \sigma_{i}^{z}\sigma_{k}^{z}~,
\end{equation}
where $\sigma_{i}^{z}$ is the $z$ Pauli matrix at site $i$. We couple the chain to two different baths at left (L) and right (R) sides, by assuming the spin-boson coupling in the $x$ component \cite{BP}. In the microscopic
approach, precisely, using the Born-Markov approximation, we obtain the LME with the dissipator $\mathcal{D} = \mathcal{D}_{L} + \mathcal{D}_{R}$, where, for $\mathcal{D}_{L}$ (similarly for $\mathcal{D}_{R}$)
\begin{align}
\Dop_L(\rhop)=&\sum_{\w>0} \lambda\w \left\{ \left[1+n_L(\w)\right] \left[\An\rhop\Ad \right.\right. \nonumber \\
& \left.- \frac{1}{2} \left(\Ad\An\rhop + \rhop\Ad\An  \right)  \right]    \nonumber \\
&+n_L(\w) \left[\Ad\rhop\An \right.   \nonumber \\
& \left. \left. - \frac{1}{2} \left(\An\Ad\rhop + \rhop\An\Ad\right)  \right]\right\}~,
\end{align}
 where $\w=\epsilon_{k}-\epsilon_{i}$ is the energy difference between the two eigenstates $|\epsilon_i\rangle$ and $|\epsilon_k\rangle$ of $H$; $n_L(\w)=[\exp(\hbar\w/\kbc T_{L})- 1]^{-1}$ is the Bose-Einstein distribution for the heat bath, and $\kbc$ is the Boltzmann constant, which we take as $1$ (such as $\hbar$). The Lindblad operator  $A_{L}(\w) = \sum_{\w} |\epsilon_i \rangle|\langle \epsilon_i |\sigma^{x}_{L}|\epsilon_k\rangle\langle \epsilon_k|$
 describes the transitions induced by the left bath. In such an approach, there is no work, and the energy flow gives indeed the heat current only \cite{FBarra}.

 The rectification phenomenon in such model is investigated in Ref.\cite{Werlang}, for the case of nearest-neighbor interaction and $N=2$, i.e., for the simpler case of a junction, not a many-body system. A regime of perfect
rectification is described. However, we need to remark that for spin models, sometimes there is a drastic difference between the junction and the many-body chains. For example, in Ref.\cite{Landi}, for the case of a homogeneous $XXZ$ model with target polarization at the boundaries and asymmetric external magnetic field, the authors show the existence of spin rectification for $N>2$, but it does not exist for $N=2$. Here, for the quantum
Ising model, if we keep only nearest-neighbor interactions, there is also a drastic difference: there are heat current and rectification for $N=2$, and no heat current for $N>2$ \cite{P2019}. Interestingly, in the presence of long range interactions capable to link the first to the last site of the chain, the heat current reappears together with the possibility of a perfect rectification if the temperature of one of the baths goes to zero \cite{P2019}.

We offer an explanation for the occurrence of rectification in a so simple spin chain. The baths, which act on all the Hamiltonian eigenstates (which in turn involve the inner sites), have an elaborate temperature dependence due to their quantum nature. In the study of the heat current, such a temperature correlation appears in the expressions for the heat current computation involving the inner sites. Thus, we have an asymmetry in the structure, e.g., a graded system, together with an
asymmetrical temperature dependence. As we invert the thermal baths, the temperature distribution changes, but not the inner structural asymmetry. Consequently the framework behind the heat current changes and thermal rectification appears. We believe that the rule played by the LRI here is only to break some hidden symmetry that prevents the on-set of heat current in many-body chains with nearest-neighbor interactions. Sometimes, due to the existence of
symmetries in the LME and in the density matrix, some spin chains have zero spin or energy currents even in the presence of strong boundary gradients \cite{PopLi}.

\section{Final Remarks}

This mini-review focuses on attempts devoted to the search of mechanisms leading to the increasement of the thermal rectification in general materials, with the aim of building efficient and feasible thermal diodes.
We describe some findings on classical and on quantum systems.

For the case of classical chains of anharmonic oscillators, the most recurrently used model for the study of heat conduction in insulating solids since Debye and Peierls, we describe the ubiquitous occurrence of heat rectification in graded materials. Moreover, we consider the study of the mechanism of heat conduction in the presence of long range
interactions, which is an interesting problem by itself. We show
that the combination of graded structures with long range interactions can really increase the rectification factor.

In the quantum regime, we present the spin chains as appropriate materials for the investigation energy rectification. We describe interesting properties for the spin and energy currents, such as the possibility of energy rectification reversal, due to manipulations in the external magnetic field.  Moreover, we show two cases of perfect diodes: one for the spin current in a two-segmented $XXZ$ model, and another for the heat current in a simple graded quantum Ising model with interaction long enough to link the first and last sites.

We are confident that such results describing interesting properties of thermal rectification in simple models and mechanisms, i.e., we may say, the description of a minimal itinerary for large rectification,   will stimulate more theoretical and also experimental research on the theme.

{\it Acknowledgments}:  Considerable part of the works recalled here was carried out with the help of different collaborators, in particular, of G. Casati. The author was partially supported by CNPq (Brazil).



\end{document}